\documentclass[aps,twocolumn,nofootinbib, superscriptaddress]{revtex4-1}
\usepackage{amssymb}
\usepackage{amsmath}
\usepackage{amstext}
\usepackage{amsfonts}
\usepackage{bbold}
\usepackage{slashed}
\usepackage{hyperref}

\def\l{\left}
\def\r{\right}

\def\d{\partial}

\def\be{\begin{eqnarray}}
\def\ee{\end{eqnarray}}

\newcommand{\sfrac}[2]{{\textstyle\frac{#1}{#2}}}

\begin{document}

\title{Spontaneously broken mass} 

\author{Solomon Endlich}
\affiliation{Institut de Th\'eorie des Ph\'enom\`enes Physiques, \\EPFL,
Lausanne, Switzerland}
\author{Alberto Nicolis}
\affiliation{Physics Department and Institute for Strings, Cosmology, and Astroparticle Physics,\\
  Columbia University, New York, NY 10027, USA}
\author{Riccardo Penco}
\affiliation{Physics Department and Institute for Strings, Cosmology, and Astroparticle Physics,\\
  Columbia University, New York, NY 10027, USA}

\begin{abstract}
The Galilei group involves {mass} as a central charge. We show that the associated superselection rule is incompatible with the observed phenomenology of superfluid helium 4: this is recovered only under the assumption that mass is {\it spontaneously broken}. This remark is somewhat immaterial for the real world, where the correct space-time symmetries are encoded by the Poincar\'e group, which has no central charge. Yet it provides an explicit example of how superselection rules can be experimentally tested. We elaborate on what conditions must be met for our ideas to be generalizable to the relativistic case of the integer/half-integer angular momentum superselection rule.
\end{abstract}

\maketitle

%%%%%%%%%%%%%%%%%%%%%%%%%%%%%%%%%%%%%%%%
%%%%%%%%%%%%%%%%%%%%%%%%%%%%%%%%%%%%%%%%

Consider a superfluid at zero temperature. For our purposes, it is useful to
characterize its dynamics in terms of its symmetry breaking pattern. The simplest superfluid is a homogeneous, isotropic thermodynamical system that has a finite density for a spontaneously broken conserved charge. In the case of superfluid helium 4, the charge in question is helium atom number. The presence of a charge density obviously breaks boosts (Lorentz's and Galilei's alike), but what is less obvious is that it also breaks time translations \cite{Nicolis:2011pv}. This is because the ground state $| \mu \rangle$ of the system minimizes the shifted Hamiltonian
\be
H' = H - \mu Q \; ,
\ee
where $H$ is the original Hamiltonian, $\mu$ the chemical potential, and $Q$ the charge. But then $| \mu \rangle$ is an eigenstate of $H'$. Given that it is not an eigenstate of $Q$---or else $Q$ would {\em not} be spontaneously broken---it cannot be an eigenstate of $H$ either. So, time translations are broken, along with $Q$, down to the diagonal combination generated by $H'$.

This pattern of symmetry breaking is all one needs to write down the low-energy effective field theory of the Goldstone excitations, the superfluid phonons. In the relativistic case this is well understood---see e.g.~\cite{Son:2002zn}. At lowest order in the derivative expansion, the effective Lagrangian is
\be
{\cal L} = P(-\d_\mu \psi \,  \d^\mu \psi) \; , \label{P(X)}
\ee
where $\psi$ is a Lorentz scalar, to be expanded as 
\be
\psi = \mu t + \pi
\ee
with $\mu$ the chemical potential, $\pi$ the phonon field, and $P$ a function that gives the equilibrium pressure as a function of the chemical potential.
Notice that one Goldstone field suffices to realize non-linearly all the broken symmetries (three boosts and the charge): this is a common phenomenon that arises when spacetime symmetries are spontaneously broken---one excitation can serve simultaneously as the Goldstone mode for several broken symmetries~\cite{Ivanov:1975zq,Low:2001bw}.  This is made systematic in the coset construction of effective field theories for Goldstones, where the possible identification of a priori different Goldstone fields goes under the name of \emph{inverse Higgs constraints}. Which inverse Higgs constraints are possible---i.e., compatible with the symmetries---is a purely algebraic question, which can be settled once the symmetry algebra and the symmetry breaking pattern are known. (On the other hand, whether the inverse Higgs constraints are actually implemented by Nature is a statement that in principle depends on the physical system under consideration~\cite{Nicolis:2013sga}.) For relativistic superfluids, one finds that, indeed, there are enough possible inverse Higgs constraints to re-express all the boost Goldstones in terms of the Goldstone associated with $Q$~\cite{Nicolis:2013lma}.

For non-relativistic superfluids in the lab, like helium 4 for instance, one expects Poincar\'e invariance to be very accurately approximated by Galilei invariance. In particular, one expects to be able to recover the low-energy dynamics of the Goldstone excitations by focusing on the Galilei group from the start. However, the mathematics of the Galilei group exhibits additional subtleties compared to that of the Poincar\'e group, which will turn out to be of great physical import in our case.

Consider a Galilei-invariant theory. The Galilei algebra is
\begin{align}
[J_i, J_j] & = i \epsilon_{ijk} \, J_k \, , \quad [J_i, K_j] = i \epsilon_{ijk} \, K_k \, , \quad [J_i, P_j] = i  \epsilon_{ijk} P_k \nonumber \\ 
[K_i, P_j]   & = -i \delta_{ij} \, M  \, , \quad [K_i,H_{\rm NR}]  = -i \,  P_i  \label{algebra}
\end{align}
and zero for all other commutators. $\vec P$, $\vec J$, and $\vec K$ generate respectively spatial translations, spatial rotations, and Galilei boosts. $H_{\rm NR}$ is the non-relativistic Hamiltonian, which, in cases where Galilean invariance is just a low-speed approximation of a more fundamental Poincar\'e invariance, is defined simply as\footnote{Here and in what follows we are setting the speed of light to one.}
\be
H_{\rm NR} = P^0 - M \; .
\ee

$M$ is the total mass of the system and, in the simplest interpretation of the Galilei group, it is just a \emph{number}---a parameter fixed once and for all. In particular, its appearance on the right hand side of the $[K,P]$ commutator forces the Galilei group to only have projective representations, with a superselection rule for the mass itself~\cite{Weinberg:1995mt}: one cannot prepare a quantum mechanical state that is a linear combination of states of different mass. We now show that superfluid helium 4 directly violates this rule.

If we interpret $M$ as a pure number, then the symmetry breaking pattern for non-relativistic superfluid helium 4 at zero temperature can be summarized as follows:
\begin{align}
\mbox{broken:} &  \left\{
\begin{array}{l}
\vec K \\
Q 
\end{array}
\right.  \label{broken} \\
& \nonumber \\
\mbox{unbroken:}&  \left\{
\begin{array}{l}
\vec P\\
\vec J \\
H' \equiv H_{\rm NR} - \mu_{\rm NR} Q.
\end{array}
\right.  \label{unbroken}
\end{align}
$Q$ generates the spontaneously broken internal $U(1)$ symmetry, and can be identified with helium atom number. $H'$ generates the unbroken time translations, which must differ from those generated by $H_{\rm NR}$, for the reason spelled out above. We are denoting the non-relativistic chemical potential by $\mu_{\rm NR}$, which is related to the relativistic one by
\be
\mu_{\rm NR} = \mu - m \; , 
\ee
where $m$ is the helium atom's mass (assuming $Q$ counts the number of helium atoms.)
Now, since we have four broken generators, we have in principle four Goldstone fields. Whether some of these are redundant and can be eliminated in favor of the others depends first of all on which inverse Higgs constraints are allowed by the symmetries. The rule of thumb is that inverse Higgs constraints are associated with non-trivial commutators of the form
\be
[T, B_1] \supset B_2 \; ,
\ee
where $T$ generates an {\em unbroken} spatial or temporal translation, and  $B_1$ and $B_2$ are broken generators. For such a commutator, there is an inverse Higgs constraint relating the Goldstone of $B_1$ to derivatives of the Goldstone  of $B_2$. For the case under consideration, the commutators of the unbroken translations with the broken generators are schematically:
\begin{align}
[P, K] & \sim M \; , \qquad [P, Q] = 0  \label{helium inverse} \\
 [H', K] & \sim P \; , \qquad [H', Q] = 0 
\end{align}
Since no right hand side involves broken generators, there are no inverse Higgs constraints that one can impose to remove some of the Goldstone excitations. We are stuck with four Goldstone modes: one for $Q$, and three for the boosts, in contradiction with the observed physical properties of helium 4.  This is to be contrasted with the relativistic analysis \cite{Nicolis:2013lma}, where the corresponding commutators are
\begin{align}
[P, K]  & \sim H = H' + \mu Q \; , \qquad  [P, Q] = 0  \\
 [H', K] & \sim P \; , \qquad \qquad  \qquad \quad [H', Q] = 0 
\end{align}
The appearance of $Q$ on the right hand side of the $[P,K]$ commutator tells us that we can re-express the Goldstones of $K$ as spatial derivatives of the Goldstone of $Q$, thus leaving us with only one independent Goldstone excitation---the phonon.

The bold conclusion at this point would be that the mere existence of superfluid helium 4 as we know it---a very non-relativistic system---{\em requires} full Poincar\'e invariance, rather than just its Galilean limit. An intriguing statement indeed. Could we have discovered special relativity this way?

There is however a more conservative and  more judicious path that one can take. As emphasized in~\cite{Weinberg:1995mt}, the interpretation of the mass as essentially a fixed parameter with a superselection rule is just one possible interpretation of the Galilei symmetry algebra. Another possibility is to add the mass to the algebra as a {\em generator} that commutes with all the others. From this perspective, the mass is now just conserved, rather than superselected. However, it is clear that if we merely add mass as an unbroken generator in \eqref{unbroken}, our algebraic analysis above still applies and we end up with four Goldstones. But there is now another possibility open to us. In the absence of a superselection rule, we can take linear combinations of mass eigenstates with different eigenvalues. That is, we can consider states that are {\em not} mass eigenstates, i.~e.~ we can {\em spontaneously break mass}. Doing so will allow us to remove three of the four Goldstones in order to correctly describe the non-relativistic superfluid state.

In hindsight, it is obvious that $M$ must be broken. Each helium atom carries mass $m$ and helium-atom number one, so that we have
\be
M = m Q 
\ee
as an operator identity, in the sense that $M$ and $m Q$ act in identical ways on the helium atoms' wavefunction\footnote{This is an artifact of our focusing on a system of just one species of particles, all with the same charge/mass ratio. See also a related discussion in~\cite{Greiter:1989qb}.}.
If $Q$ is spontaneously broken, so is $M$.

Whenever we encounter $M$ in our analysis, we can now replace it with $mQ$. In particular, in the $[P,K]$ commutator of eq.~\eqref{helium inverse} we now have $Q$ on the right hand side, which is spontaneously broken. We can thus eliminate the three Goldstones associated with $\vec K$ in favor of that associated with $Q$, as desired. 

As a more quantitative check  that this is indeed the correct description of a non-relativistic superfluid like helium 4, we now carry out explicitly the coset construction of the Goldstone low-energy effective field theory.
Following~\cite{Volkov:1973vd,ogievetsky:1974ab}, we parameterize the coset for the symmetry breaking pattern \eqref{broken}, \eqref{unbroken} as
\be
\Omega(x) = e^{-i H't + i   \vec P \cdot \vec x} e^{i \pi(x) Q} e ^{i \vec \eta(x) \cdot \vec K}  \; ,
\ee
where $\pi(x)$ and $\vec \eta(x)$ are the four Goldstone fields.
Using recursively the Galilei algebra  \eqref{algebra} (with $M$ replaced by $mQ$), we find the Maurer-Cartan form $\Omega^{-1} d \Omega$:
\begin{align}
\Omega^{-1} \d_t \Omega & = i \big[-H' + \vec \eta \cdot \vec P + \d_t \vec \eta \cdot \vec K + (\d_t \pi - \sfrac m 2 \eta^2)Q \big] \\
\Omega^{-1} \d_j \Omega & = i \big[ P_j + \d_j \vec \eta \cdot \vec K + (\d_j \pi - m \, \eta_j)Q \big]
\end{align}
These expressions can be conveniently combined using a standard relativistic notation,
\be
\Omega^{-1} \d_\mu \Omega = i e_\mu{}^\alpha ( \bar{P}_\alpha + \nabla_\alpha \pi \, Q + \nabla_\alpha \eta^j \,K_j )
\ee
where $\bar P_\alpha = (-H', \vec P)$ is the four-vector of unbroken translations, whose cofficients
\be
e_\mu{}^\alpha = \delta_\mu^0 \delta^\alpha_0 + (\delta_\mu^j  + \delta_\mu^0 \eta^j ) \delta^\alpha_j
\ee
play the role of a tetrad, and the covariant derivatives of the Goldstone fields are
\begin{align}
\nabla_\alpha \pi &= \delta_\alpha^0 \l[ \d_0 \pi +\sfrac{m}{2} \eta^2 - \eta^i \d_i \pi \r] + \delta_\alpha^j (\d_j \pi - m \,\eta_j ) \\
\nabla_\alpha \eta^i &= \d_\alpha \eta^i + \mbox{higher orders}
\end{align}
The appearance of an undifferentiated $\vec \eta$ in the spatial covariant derivatives of $\pi$,
\be
\nabla_j \pi = \d_j \pi - m \,\eta_j  \; , \label{inverse Higgs}
\ee
tells us that we can set these to zero and re-express the boost Goldstones $\vec \eta$ in terms of derivatives of $\pi$:
\be
\eta_j = \sfrac1m \d_j \pi \; .
\ee
These are the three inverse Higgs constraints alluded to above. Once we do that, we see that the only covariant derivatives left are
\be
\nabla_0 \pi =  \d_0 \pi -\sfrac{1}{2m} (\d_i \pi \d_i \pi) 
\ee
and $\nabla_\alpha \vec \eta$. Upon using \eqref{inverse Higgs}, the latter involves second derivatives of $\pi$, and can thus be ignored to lowest order in the derivative expansion. 
We reach the conclusion that to lowest order, the low-energy effective action for our superfluid must take the form
\footnote{The measure should be multiplied by the determinant of the tetrad $e_\mu {}^ \alpha$, which in our case is simply one.}
\be
S = \int d^3 x dt \,  P\big(\d_0 \pi -\sfrac{1}{2m} (\d_i \pi \d_i \pi) \big) \; ,
\ee
which is indeed the correct description of a superfluid obeying Galilean relativity~\cite{Greiter:1989qb}. Notice that this is also what one gets by taking directly the non-relativistic limit  of the relativistic action \eqref{P(X)} and, for consistency, neglecting $\mu_{\rm NR}$ compared to $m$.

Our deconstructing these subtle aspects of the Galilei algebra may appear academic, since it is the Poincar\'e group that---as far as we know---correctly describes the spacetime symmetries of Nature. However, we find our Galilei algebra example instructive, because it identifies spontaneous symmetry breaking as a potentially generic manifestation of violations of superselection rules. 

For the Poincar\'e group, there are no central charges in the algebra, but there is a superselection rule of topological origin~\cite{Weinberg:1995mt}: it states that it is not possible to prepare linear superpositions of states of integer {\em and} half-integer angular momentum. The superselection rule disappears if one replaces the Lorentz group ($SL(2,C)/Z_2$) with its universal covering ($SL(2,C)$). Can we use spontaneous symmetry breaking to probe whether Nature respects this superselection rule?

One way to violate this superselection rule via spontaneous symmetry breaking, is to break angular momentum via the expectation value of a {\em fermionic} operator. To see why that's the case, consider the expectation value
\be
\langle \psi | {\cal O}(x) | \psi \rangle \; ,
\ee
where ${\cal O}$ is a local operator of half-integer spin---say $\sfrac12$, for simplicity.
If we expand the state $| \psi \rangle$ in angular momentum eigenstates,
\be
\label{expanded states}
| \psi \rangle = c_0 | 0 \rangle + c_{\frac12} | \sfrac12 \rangle + c_{1} | 1 \rangle + \dots \; , 
\ee
we see that  the standard (i.e., not `super') selection rules for angular momentum imply that all non-vanishing contributions to the expectation value above, can only come from matrix elements in which the  bra's and the ket's angular momenta differ by $\sfrac12$. That is to say, for the expectation value above to be nonzero, $| \psi \rangle$ {must} contain {\em both} integer {\em  and} half-integer spins. A fermionic expectation value would violate the superselection rule in question.

Nonzero expectation values for fermionic operators are usually dismissed as impossible, on the basis of two independent arguments:
\begin{enumerate}
\item
They would break Lorentz invariance (this applies to all non-scalar operators, fermionic and bosonic alike);
\item
They would violate Pauli-exclusion principle: occupation numbers can be either zero or one, leaving no room for `large',  `classical' backgrounds.
\end{enumerate}
However, as T.~D.~Lee once remarked, if one feels that two arguments are needed to explain one  phenomenon, one should become suspicious \cite{Weinberg_talk, TDL}.
The first objection is clearly irrelevant for ordinary condensed matter systems, which break Lorentz boosts by selecting a preferred rest frame. Notice however that in addition to boosts, a fermionic expectation value necessarily breaks rotations as well. This is to be contrasted with bosonic operators, which can have Lorentz-violating {\em isotropic} expectation values, like for instance $\langle V^\mu \rangle = \delta^\mu_0$. Still, there is no shortage of anisotropic condensed matter systems. From the symmetry viewpoint alone, these could in principle harbor a fermionic expectation value.

The second objection relies on a perturbative idea of fermionic operators---as fields that destroy and create Fock states. Beyond perturbation theory, for a strongly coupled system for instance, we do not see any obstruction of this kind to a fermionic expectation value. 
For instance, do we know for a fact that QCD does not admit a Lorentz-violating, anisotropic phase with a fermionic expectation value $\langle {\cal O} \rangle \sim \Lambda^\Delta_{\rm QCD}$, where $\Delta$ is the dimension of ${\cal O}$? 
Also, there is no need for such expectation values to be `large' or `classical' in any sense. They could be of order one in the natural units of the system under consideration, and they would still violate our superselection rule. Consider for instance the theory of a free massive Dirac fermion, in a superselection rule-violating state
\be
|\psi \rangle = | 0 \rangle + | \vec p =0 , + \rangle \; , 
\ee
where $| 0 \rangle$ denotes the theory's vacuum, and $| \vec p =0 , + \rangle$ denotes a single fermion at rest, polarized in the positive $z$ direction. The expectation value of the field operator $\Psi(x)$ that annihilates the fermion is nonzero:
\be \label{free case}
\langle \psi | \Psi(x) | \psi \rangle \propto u^+_{\vec p = 0} e^{-imt} \; ,
\ee
where $u^+_{\vec p = 0} $ is the `coefficient function' for a spin-up fermion at rest:
\be
u^+_{\vec p = 0} = \sqrt{m} (1,0,1,0) \; .
\ee
Note the time dependence in eq.~(\ref{free case}). As we can already see from this toy example, such non-zero fermionic expectation value states will generically break time translations. That is because states of the form eq.~(\ref{expanded states}) are generically not energy eigenstates. This is, of course, nothing to be alarmed by as the same thing occurs in ordinary superfluids as discussed above.

Also notice that there is nothing Grassmanian about the right hand side of \eqref{free case}. One might have thought
that a nonzero  expectation value for a fermionic operator {\em must} be a Grassman variable, since at first sight it should be a number that---like the operator itself---anti-commutes with any other Grassman variable $\theta$.
However, this `proof' implicitly uses the idea that such a $\theta$ {\it either} commutes {\it or} anticommutes with the state $|\psi \rangle$ of the system, depending on the total spin of $|\psi \rangle$. But, upon closer inspection, it is clear that for states $| \psi \rangle$ that violate our superselection rule, like the state above for instance, $\theta$ {\it neither} commutes {\em nor} anticommutes with $| \psi \rangle$, simply because $| \psi \rangle $ does not have a definite spin parity:
\be
\theta \big ( | 0 \rangle + | \vec p =0 , + \rangle \big) = \big ( | 0 \rangle - | \vec p =0 , + \rangle \big) \theta \; .
\ee

This invalidates the argument, and in fact makes it clear that in general a nonzero fermionic expectation value will {\it always} turn out to be a set of ordinary numbers: as noticed above, it can only get contributions  from mixed boson-fermion matrix elements of the fermionic operator in question, which {\it commute} with all Grassman variables. 

This does not necessarily mean that fermionic expectation values would be directly measurable by a macroscopic probe, like for instance, an electric field is. Yet, in other situations we attach great physical significance to entities that can  be experimentally accessed only indirectly---like the wave function of a quantum system for instance. We see no a priori reason why fermionic expectation values cannot fall in the same category. Consider for instance a Yukawa-type interaction, coupling a fermion bilinear to a scalar. In a state in which one of the {\it fermions} gets a vev, fermion number is not a good quantum number for excitations anymore, and one can have a nonzero amplitude for boson to fermion conversion processes, which can be  tested experimentally.

We are thus left with two challenging questions which we leave for future work: 
\begin{enumerate}
\item Can we devise an anisotropic condensed matter system that has a chance of exhibiting a nonzero expectation value for a fermionic operator?
\item How do we experimentally test whether such an expectation value is actually there, that is, what would its observable physical consequences be?
\end{enumerate}

\vspace{.5cm}

{\it Acknoledgements---}
We would like to thank Nima Arkani-Hamed, Sergei Dubovsky, Lam Hui, Dan Kabat, and Rachel Rosen for stimulating discussions. This work was partially supported by NASA under contract NNX10AH14G and by the DOE under contract DE-FG02-11ER41743.

%%%%%%%%%%%%%%%%%%%%%%%%%%%%%%%%%%%%%%%%
%%%%%%%%%%%%%%%%%%%%%%%%%%%%%%%%%%%%%%%%

%\vspace{1cm}
%
%\noindent {\bf Acknowledgments:}   This work was supported by NASA under contract NNX10AH14G and by the DOE under contract DE-FG02-11ER41743.

\bibliographystyle{apsrev4-1}
\bibliography{biblio}

\end{document}